\documentclass[aps,prx,reprint,superscriptaddress,nofootinbib]{revtex4-2}

\usepackage{amsmath,amssymb,amsfonts,bm}
\usepackage{booktabs,array}
\usepackage{graphicx}
\graphicspath{{./}{Figures/}}
\usepackage{hyperref}
\usepackage{xurl}
\usepackage{xcolor}
\usepackage{mdframed}
\usepackage{microtype}
\usepackage{tikz}
\usetikzlibrary{arrows.meta,positioning,fit,backgrounds,calc}

\hypersetup{
    colorlinks=true,
    linkcolor=blue!60!black,
    citecolor=blue!60!black,
    urlcolor=blue!60!black
}

\newcommand{\supp}{\operatorname{supp}}
\newcommand{\MI}{\operatorname{I}}
\newcommand{\DC}{\operatorname{DC}}
\newcommand{\EXmean}{E_{X,\mathrm{mean}}}

\begin{document}

\title{Observable-Conditioned Backaction in Dynamic Circuits: A Higher-Order Context-Conditioned Kernel for Local Dynamics}

\author{Petr Sramek}
\email{p.sramek@whytics.com}
\affiliation{Whytics, Cambridge, MA, USA}
\affiliation{DAGI Research Program, Foundations Track}

\date{March 2026}

\begin{abstract}
Mid-circuit measurements are essential primitives for dynamic circuits and quantum error correction, yet the scalable characterization of their induced disturbance on spectator qubits remains a central practical problem. Modern open-quantum-systems theory already treats such disturbance through quantum instruments and multitime process descriptions, but device-level benchmarking often compresses it into low-order proxy metrics such as $T_1$, $T_2$, readout assignment error, and pairwise crosstalk matrices. In this work we argue that these proxies can be operationally incomplete for multiscale dynamic circuits.

We therefore introduce a \emph{higher-order context-conditioned kernel}, \(\Gamma_{\rm eff}[Y,\mathcal{O}] = \Gamma_{\rm loc}[\mathcal{O}] + \Gamma_{\rm proxy}[\mathcal{O}] + \Gamma_{\rm rel}[Y,\mathcal{O}]\), where $Y$ denotes a global context label and $\mathcal{O}$ the queried local observable. The new term $\Gamma_{\rm rel}[Y,\mathcal{O}]$ is presented not as a fully predictive microscopic law, but as a phenomenological compression ansatz that isolates residual context dependence not explained by standard low-order proxies. To avoid the known impossibility issues of quantum partial-information decompositions on non-commuting operator algebras, the M\"obius weights entering this ansatz are evaluated operationally on \emph{classical} measurement outcomes.

The paper proceeds in three steps. First, we use earlier GHZ-versus-clock hardware results as empirical motivation for an observable-class split. Second, we present the primary dynamical evidence using the A6 harness. Crucially, A6 is a \emph{synthetic hardware harness}: it uses a programmed conditional interaction to inject a known, pure higher-order context dependence. Because the $(C_0,C_1,C_2)$ parity context is invisible to singles and pairs by construction, we demonstrate that standard low-order device diagnostics are fundamentally blind to the source of the probe's disturbance. Third, we demonstrate coherent controllability through the A6.2 quantum-eraser experiment, where programmable MARK interactions suppress unconditional fringes while eraser-basis conditioning restores them, consistent with complementarity bounds. Taken together, these results validate a context-conditioned description of backaction over proxy-only null models and set the stage for future discovery-mode scans of unprogrammed parasitic crosstalk. To isolate this graph-theoretic core from hardware-specific noise, we also analyze a noiseless arithmetic control model: the squarefree divisor / log-prime quota complex of the integers. Because the prime-number network has no microwave stack, relaxation channel, or SPAM layer, it provides a theorem-level benchmark. In that exact setting, rigid quotas strand an $x/\log x$-scale floor of isolated 1-body prime vertices, while softened logistic boundaries restore parity-resolved cancellation, furnishing a noiseless analogue of the hard-cut versus soft-cut logic observed on hardware.
\end{abstract}

\keywords{dynamic quantum circuits, mid-circuit measurement, spectator-qubit backaction, higher-order correlations, superconducting qubits, quantum characterization}

\maketitle

\section{Introduction}
\label{sec:intro}

Dynamic quantum circuits require mid-circuit measurements, resets, and feed-forward control to act on some qubits while preserving useful coherence on others. This is now a standard engineering problem in quantum error correction, dynamic compilation, and measurement-based subroutines. In the formal language of open quantum systems, the disturbance induced by such operations is not assumed to be a featureless scalar field; it is described by quantum instruments, multitime processes, and process-tensor style formalisms when one seeks full microscopic detail \cite{Pollock2018,Milz2021,Rudinger2022}. In practice, however, scalable benchmarking rarely retains that full structure. Device characterization frequently collapses disturbance into low-order proxy metrics such as $T_1$, $T_2$, readout assignment error, residual cavity population, or pairwise crosstalk maps \cite{Rudinger2022,Govia2023,Hothem2025}.

Those proxy models are often excellent engineering summaries, but they are not guaranteed to be sufficient summaries for every queried observable class. In particular, they may fail when the experimentally relevant question is not ``what happens to an isolated qubit?'' but rather ``what happens to a local probe conditioned on a higher-order context that is deliberately hidden from all 1- and 2-body summaries?'' The central claim of this paper is that such a regime can be built and diagnosed on present superconducting hardware, and that its natural operational summary is not a single scalar backaction number but a decomposition into local, proxy, and residual context-conditioned terms.

The strategic role of this paper is therefore focused and pragmatic. We do not claim to replace open-system theory. We instead propose a practical compression ansatz for a specific characterization gap: dynamic-circuit backaction that depends on the \emph{class} of context and observable being queried, even after the usual low-order hardware explanations are explicitly controlled. Our primary empirical target is the A6 harness. Earlier GHZ-versus-clock contrasts remain useful as motivation, but by themselves they leave room for standard objections about state fragility, embedding differences, and correlated readout artifacts.

The architecture of the paper is intentionally conservative. Section~\ref{sec:claims} states the operational definitions and the status of claims. Section~\ref{sec:motivation} uses earlier T4/C1 results only to motivate an observable-class split. Section~\ref{sec:a6} then presents the real dynamical core: the A6 parity-context experiment on \texttt{ibm\_boston}, with its corrected local witness, parity balancing, passive nulls, CTXONLY screens, and permutation tests \cite{SramekA6Context2026}. Section~\ref{sec:a62} treats coherent controllability via the dedicated A6.2 MARK/ERASE experiment, framed explicitly as a hardware-native quantum eraser rather than as unconditional macroscopic revival \cite{SramekA62Erasure2026}. We conclude with a restrained discussion of what this framework establishes, what it does not, and which scaling studies would be required before stronger predictive claims are justified.

To further separate structural graph effects from any suspicion of microwave bleed, spectator relaxation, or readout artifacts, we also introduce a noiseless arithmetic control model: the squarefree divisor / log-prime quota complex of the integers \cite{Bjorner2011,Pakianathan2013}. This arithmetic construction is not a decorative analogy. It serves as a theorem-level benchmark on an infinite, exactly evaluable graph with no hardware degrees of freedom. In that setting, rigid quotas strand an $x/(2\log x)$-scale floor of isolated prime vertices, while softened Fermi--Dirac/$\tanh$ boundaries continuously heal the shell and strengthen parity-resolved cancellation. The role of the arithmetic section is therefore precise: it isolates the combinatorics of graph cuts from the separate QCVV problem of which physical channel realizes them on superconducting hardware.

\section{Status of Claims and Operational Definitions}
\label{sec:claims}

To separate operational claims from interpretive ones, and to keep the engineering content insulated from broader DAGI language, we state the core definitions explicitly.

\begin{mdframed}[linecolor=black, linewidth=1pt, roundcorner=3pt, innertopmargin=12pt, innerbottommargin=12pt, innerleftmargin=15pt, innerrightmargin=15pt, backgroundcolor=gray!5]
\textbf{\large BOX 1: Status of Claims and Operational Definitions}
\vspace{0.5em}\hrule\vspace{0.8em}

\textbf{1. Definition: Higher-Order Context-Conditioned Kernel.}\\
Let $Y$ denote a context label and $\mathcal{O}$ an observable or witness queried on a dynamic circuit. We write the total effective backaction as
\begin{equation}
\Gamma_{\rm eff}[Y,\mathcal{O}] =
\Gamma_{\rm loc}[\mathcal{O}] +
\Gamma_{\rm proxy}[\mathcal{O}] +
\Gamma_{\rm rel}[Y,\mathcal{O}],
\label{eq:gammaeff}
\end{equation}
and, when useful, define an associated suppression factor
\begin{equation}
D[Y,\mathcal{O}] \equiv \exp\!\big(-\Gamma_{\rm eff}[Y,\mathcal{O}]\big).
\label{eq:Ddef}
\end{equation}
Here $\Gamma_{\rm loc}$ collects ordinary local decoherence channels of the probe, while $\Gamma_{\rm proxy}$ collects standard low-order hardware mechanisms such as spectator dephasing, residual cavity photons, microwave bleed, and pairwise couplings. The residual term is modeled as
\begin{equation}
\Gamma_{\rm rel}[Y,\mathcal{O}] =
\sum_{v\in \mathcal{R}(Y)} r_v
\sum_{\substack{S \subseteq \supp(Y)\\ |S|\ge 3}}
\tilde f^{+}_{Y}(S)\,\omega_v(S,\mathcal{O}),
\label{eq:gammarel}
\end{equation}
where $\mathcal{R}(Y)$ is the set of record-producing or context-defining events, $r_v$ is an operational record weight, $\tilde f^{+}_{Y}(S)$ is a normalized positive M\"obius weight for subset $S$, and $\omega_v(S,\mathcal{O})\in[0,1]$ is an operational overlap between event $v$, the higher-order context subset $S$, and the queried observable $\mathcal{O}$.

The term ``M\"obius weight'' is meant operationally here: to avoid the known impossibility results for unique operator-algebraic partial information decompositions on non-commuting quantum states, the present paper evaluates the $f^{+}$-type quantities on the \emph{Shannon entropy of classical measurement outcomes}, not on non-commuting density operators directly.

\vspace{0.8em}
\textbf{2. Operational Proposition: Context-Dependent Local Dynamics.}\\
Let $Y$ be a context variable carried by a context register $C=\{C_0,C_1,\dots\}$ and let $\mathcal{O}$ be a witness on a probe subsystem $P$. We say that context-dependent local dynamics, denoted $\DC(Y\to \mathcal{O})$, holds operationally if
\begin{equation}
\MI(Y;C_i)\approx 0,\qquad \MI(Y;C_iC_j)\approx 0
\quad \forall\ i,j,
\label{eq:dcdef}
\end{equation}
while the effective dynamics of $\mathcal{O}$ nevertheless depend measurably on $Y$ under matched local controls. In words: the context is engineered to be invisible to all 1- and 2-body summaries of its carrier set, but the probe still tracks that context.

\vspace{0.8em}
\textbf{3. Empirical Claim Evaluated Here.}\\
The A6 experiment family provides evidence that a proxy-only model,
\[
\Gamma_{\rm eff}[Y,\mathcal{O}] \approx
\Gamma_{\rm loc}[\mathcal{O}] + \Gamma_{\rm proxy}[\mathcal{O}],
\]
is operationally incomplete for the tested circuits. The present evidence supports this claim in the limited sense of a hardware harness with engineered contexts, strong passive controls, lane balancing, and coherent eraser variants. It is not yet a claim about arbitrary unprogrammed many-body environments.

\vspace{0.8em}
\textbf{4. Interpretive Mapping.}\\
Within DAGI, $\Gamma_{\rm rel}[Y,\mathcal{O}]$ is interpreted as a structured record-sensitive term associated with how a queried observable intersects the context-defining history graph. Nothing in the empirical sections requires that interpretation: the operational content of the paper is compatible with standard quantum instruments, decoherence theory, and the textbook quantum-eraser mechanism.
\end{mdframed}

\subsection{A noiseless arithmetic analogue of hard cuts and soft cuts}

The residual kernel in Eq.~(\ref{eq:gammarel}) is motivated experimentally, but the graph-cut logic it summarizes can also be tested in a setting with no microwave control stack, no $T_1/T_2$ channels, and no SPAM layer at all. A convenient exact model is the squarefree divisor complex
\begin{equation}
\Delta_x \equiv \left\{S\subseteq \mathbb{P}\,:\, \prod_{p\in S} p \le x\right\},
\label{eq:Deltax}
\end{equation}
where $\mathbb{P}$ is the set of primes and the empty face is included. This is simultaneously Bj\"orner's simplicial complex of squarefree integers up to $x$ and the log-prime quota complex with vertex weights $w(p)=\log p$ and quota $q=\log x$ \cite{Bjorner2011,Pakianathan2013}. The two descriptions are equivalent because
\[
\prod_{p\in S} p \le x
\quad\Longleftrightarrow\quad
\sum_{p\in S}\log p \le \log x.
\]

In this arithmetic model the reduced Euler characteristic is exactly the Mertens sum,
\begin{equation}
\tilde\chi(\Delta_x)
=
-\!\!\sum_{n\le x}\mu(n)
=
-M(x),
\label{eq:mertens-euler}
\end{equation}
so odd and even prime-factor layers alternate with the same M\"obius sign structure that underlies the inclusion--exclusion weights used elsewhere in this paper \cite{Bjorner2011}. The advantage of Eq.~(\ref{eq:mertens-euler}) is not that it proves a new statement about the primes; it is that it gives a completely noiseless, exactly evaluable graph in which one can test the same boundary-cut logic invoked operationally in A6 and A6.2.

A direct lemma already captures the key hard-cut effect.

\paragraph*{Lemma (isolated prime shell).}
If $p$ is a prime with $x/2 < p \le x$, then $p$ is an isolated vertex of $\Delta_x$.

\paragraph*{Proof.}
Any 1-simplex containing $p$ would have the form $\{p,q\}$ with $q$ another prime and hence $q\ge 2$. But then $pq \ge 2p > x$, so $\{p,q\}\notin \Delta_x$. Therefore $p$ has no incident edges and is isolated. \hfill $\square$

Thus the hard quota creates an exact floor of
\begin{equation}
N_1(x)=\pi(x)-\pi(x/2)
\sim \frac{x}{2\log x}
\qquad (x\to\infty),
\label{eq:isolated-prime-floor}
\end{equation}
isolated 1-body defects, by the prime number theorem. In the operational dictionary used throughout this paper, the hard mathematical quota $n\le x$ is the exact arithmetic analogue of a macroscopic \texttt{REC} boundary. The large singleton primes $p\in(x/2,x]$ carry interaction order $k=1$: any attempt to extend them by another prime immediately violates the quota, since $pq>x$. They therefore slip past the boundary unentangled and accumulate as a macroscopic floor of isolated defects. This is the arithmetic counterpart of Wigner's isolated local clock or spectator qubit: a $k=1$ probe that remains in unitary isolation while higher-order webs are the ones actually intersected and severed by the boundary cut. The key point is not metaphorical. It is a theorem-level example showing that hard graph cuts can be intrinsically observable-selective even on a completely noiseless causal graph.

The same quota-complex formalism also identifies where the nontrivial topology lives. For a scalar quota complex, the homotopy is carried by shell faces whose total weight lies in the interval $[q-w_{\min},q)$, where $w_{\min}$ is the smallest vertex weight \cite{Pakianathan2013}. Here $w_{\min}=\log 2$, so the arithmetic shell is
\begin{equation}
\begin{aligned}
\log x-\log 2 &\le \sum_{p\in S}\log p < \log x,\\
&\Longleftrightarrow\quad x/2 \le \prod_{p\in S} p < x.
\end{aligned}
\label{eq:arithmetic-shell}
\end{equation}
Equation~(\ref{eq:arithmetic-shell}) is the precise analogue of a graph cut localized near a record boundary: the interesting topology is concentrated in a thin shell adjacent to the cutoff, while the primes in $(x/2,x]$ are stranded as isolated vertices outside any higher-order extension.

A mathematically exact softened version is obtained by replacing the hard indicator $\mathbf{1}_{n<x}$ with the logistic / $\tanh$ profile
\begin{equation}
\begin{aligned}
W_{\tau,x}(n)
&\equiv
\frac12\!\left[1-\tanh\!\left(\frac{\log n-\log x}{\tau}\right)\right]\\
&=
\frac{1}{1+(n/x)^{2/\tau}},
\qquad 0<\tau<2.
\end{aligned}
\label{eq:softweight}
\end{equation}
Define the corresponding smoothed Euler sum
\begin{equation}
\tilde\chi_{\tau}(x)
\equiv
-\sum_{n=1}^{\infty}\mu(n)\,W_{\tau,x}(n).
\label{eq:smoothed-euler}
\end{equation}
Because $W_{\tau,x}(n)=O(n^{-2/\tau})$ as $n\to\infty$, the series in Eq.~(\ref{eq:smoothed-euler}) converges absolutely for every fixed $0<\tau<2$. Moreover, for any noninteger $x>1$,
\begin{equation}
\lim_{\tau\to 0^{+}} \tilde\chi_{\tau}(x)
=
-\sum_{n<x}\mu(n),
\label{eq:softlimit}
\end{equation}
by dominated convergence. The physical role of this soft profile is precise. The logistic/$\tanh$ weight is the arithmetic analogue of an \texttt{ERASE} operator: it does not alter the underlying multiplicative graph, but continuously softens the discontinuous shell created by the hard quota. In our numerical parity-resolved evaluations of this exact model, we find that the hard cutoff leaves phase-cancellation between the dominant odd (synergistic) and even (redundant) sectors bottlenecked at about $81.4\%$ over the computed ranges, precisely because the stranded $k=1$ prime shell cannot rejoin the composite network. Replacing the cliff by the continuous $\tanh$ boundary raises the mean cancellation efficiency to about $93.7\%$ and drives the dominant parity sectors to essentially perfect anti-correlation. In this exact arithmetic sense, \texttt{ERASE} is not the resurrection of lost microscopic dynamics; it is the programmable softening of an artificially rigid boundary, which restores access to interference patterns hidden by the hard partition.

\section{Empirical Motivation: The Observable-Class Split}
\label{sec:motivation}

Earlier DAGI hardware papers identified a striking contrast between two remotely queried observable classes: a local single-qubit clock and a distributed GHZ-like witness subjected to schedule-matched remote interventions \cite{SramekWignersFriend2026,SramekTimeDilation2026,SramekRecordErasure2026}. The important lesson for the present paper is modest. Those data motivate the suspicion that a single scalar summary of backaction is not always enough. They do \emph{not} by themselves settle the dynamical mechanism, because critics can reasonably point out that GHZ states are intrinsically fragile and that the compared circuits differ in scale and embedding.

Accordingly, the T4/C1 contrast serves here only as motivation for the type of model we should seek: one in which local/proxy channels and higher-order context-conditioned channels are separated rather than compressed into one number.

The right next experiment is therefore not another fragile global witness, but a controlled harness in which the higher-order structure lives in the \emph{context} while the measured probe remains local and robust. That is exactly what A6 was designed to supply.

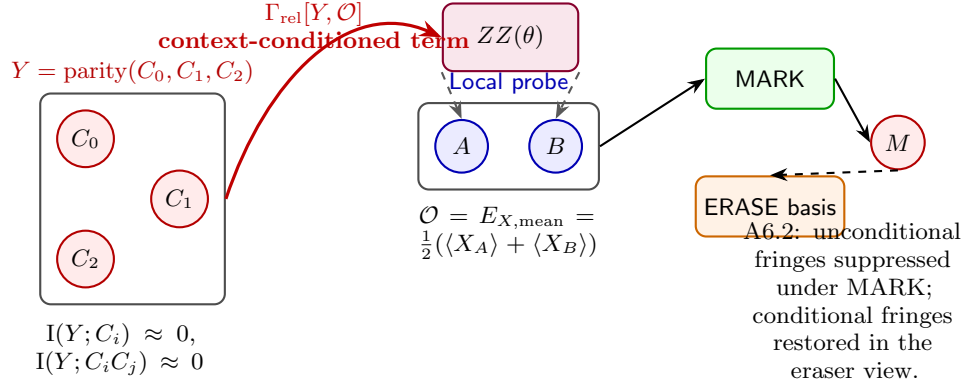
\begin{figure*}[t]
\centering
\begin{tikzpicture}[
>=Stealth,
font=\sffamily,
ctx/.style={circle,draw=red!70!black,fill=red!8,thick,minimum size=7mm},
probe/.style={circle,draw=blue!70!black,fill=blue!8,thick,minimum size=7mm},
box/.style={draw=black!70,rounded corners,thick,inner sep=6pt},
arrowrel/.style={->,very thick,red!75!black},
arrowproxy/.style={->,thick,dashed,gray!70!black},
note/.style={align=center,font=\small}
]
\node[ctx] (c0) at (0,0.9) {$C_0$};
\node[ctx] (c1) at (1.3,0.0) {$C_1$};
\node[ctx] (c2) at (0,-0.9) {$C_2$};
\node[box,fit=(c0)(c1)(c2),label={[red!70!black]above:$Y=\mathrm{parity}(C_0,C_1,C_2)$}] (ctxbox) {};
\node[note,text width=3.0cm] at (0.5,-2.05) {$\MI(Y;C_i)\approx 0$,\\ $\MI(Y;C_iC_j)\approx 0$};

\node[probe] (a) at (5.2,0.75) {$A$};
\node[probe] (b) at (6.55,0.75) {$B$};
\node[box,fit=(a)(b),label={[blue!70!black]above:Local probe}] (probebox) {};
\node[note,text width=3.1cm] at (5.9,-0.55) {$\mathcal{O}=\EXmean$\\ $=\frac12(\langle X_A\rangle+\langle X_B\rangle)$};

\node[draw=purple!70!black,fill=purple!10,rounded corners,thick,minimum width=1.9cm,minimum height=0.9cm] (zz) at (5.9,2.15) {$ZZ(\theta)$};
\draw[arrowproxy] (zz.south west) -- (a.north);
\draw[arrowproxy] (zz.south east) -- (b.north);

\draw[arrowrel] (ctxbox.east) .. controls (2.55,1.85) and (3.75,2.8) .. node[above,align=center,font=\small\bfseries,yshift=2pt] {$\Gamma_{\rm rel}[Y,\mathcal{O}]$\\context-conditioned term} (zz.west);

\node[draw=green!60!black,fill=green!8,rounded corners,thick,minimum width=1.75cm,minimum height=0.8cm] (mark) at (9.45,1.55) {MARK};
\node[draw=orange!80!black,fill=orange!10,rounded corners,thick,minimum width=1.95cm,minimum height=0.8cm] (erase) at (9.45,-0.15) {ERASE basis};
\node[ctx] (m) at (11.15,0.7) {$M$};
\draw[->,thick] (probebox.east) -- (mark.west);
\draw[->,thick] (mark.east) -- (m.west);
\draw[->,thick,dashed] (m.south) -- (erase.east);
\node[note,text width=3.45cm] at (10.05,-1.45) {A6.2: unconditional fringes suppressed under MARK; conditional fringes restored in the eraser view.};
\end{tikzpicture}
\caption{\textbf{Updated conceptual geometry of the paper.} The higher-order structure lives in the \emph{context} $Y$, not in the size of the local probe itself. In A6, the parity label on $(C_0,C_1,C_2)$ is pairwise matched by construction, while the measured probe is the local witness $\EXmean$ on $(A,B)$. In A6.2, a separate marker $M$ is used to study programmable which-path tagging and conditional interference restoration.}
\label{fig:concept}
\end{figure*}

\section{Primary Dynamical Evidence: The A6 Context Harness}
\label{sec:a6}

\subsection{Architecture and design logic}

A6 is the core dynamical experiment in this paper. It was executed on IBM \texttt{ibm\_boston} and implemented in eight repeated five-qubit lanes in order to maximize information per billed quantum second \cite{SramekA6Context2026}. Each lane uses the motif
\[
\{A,B,C_0,C_1,C_2\},
\]
where $(A,B)$ is the probe pair and $(C_0,C_1,C_2)$ carries the context. The context label
\[
Y = C_0 \oplus C_1 \oplus C_2
\]
is engineered using a pairwise-matched parity ensemble: in the ideal distribution, every single context bit and every context pair is uninformative about $Y$, while the full triple determines it \cite{SramekA6Context2026}. This is the sharpest operational way to ensure that any measured influence of $Y$ on the probe cannot be attributed to low-order context summaries alone.

The active mechanism is equally explicit. A6 computes the context parity into a control bit (implemented minimally through $C_2$ in the hardware design), applies a conditional $ZZ(\theta)$ interaction on $(A,B)$, uncomputes the parity logic, and finally measures the outputs \cite{SramekA6Context2026}.

We emphasize a crucial epistemological point regarding the programmed nature of this harness. Because the conditional $ZZ(\theta)$ gate is explicitly specified, A6 does not claim to ``discover'' an uncontrolled, natural hardware backaction channel. Instead, A6 operates as a \emph{calibrated metrological wind tunnel} for Quantum Characterization, Verification, and Validation (QCVV). It is designed to answer a specific operational question: If a multiqubit circuit contains a pure, higher-order context-dependent channel (whether due to intentional dynamic feed-forward logic or parasitic measurement crosstalk), how do standard device-level proxies interpret it? By construction, the context register's 1- and 2-body marginals remain completely mixed. Consequently, any proxy-only diagnostic (such as randomized benchmarking or pairwise crosstalk maps) is fundamentally blind to the contextual trigger and must misdiagnose the probe's suppression as featureless local noise (for example, ordinary dephasing). Thus, while the $\Delta E$ response confirms that the harness works as programmed, its primary scientific purpose is to prove that proxy models systematically fail to identify higher-order context, thereby validating our kernel against a known injected ground truth.

\subsection{Correct local witness and ideal functional form}

A6v1 corrected an important observable mistake. If the probe pair begins in $\lvert +\rangle_A\lvert +\rangle_B$ and undergoes
\[
U(\theta)=e^{-i\theta (Z\otimes Z)/2},
\]
then the two-qubit correlator $\langle X_A X_B\rangle$ is \emph{invariant} because $[Z\otimes Z,\, X\otimes X]=0$ \cite{SramekA6Context2026}. Consequently, the primary local dynamics witness is not a two-body $XX$ correlator but the single-qubit average
\begin{equation}
\EXmean \equiv \frac12\big(\langle X_A\rangle+\langle X_B\rangle\big).
\label{eq:EXmean}
\end{equation}
By Pauli conjugation one finds
\[
U^\dagger(\theta)(X\otimes I)U(\theta)
=
(X\otimes I)\cos\theta + (Y\otimes Z)\sin\theta,
\]
and similarly for $I\otimes X$. On $\lvert ++\rangle$ the mixed term vanishes, so the ideal ``interaction-on'' branch gives $\EXmean=\cos\theta$, whereas the ``interaction-off'' branch gives $\EXmean=1$ \cite{SramekA6Context2026}. A natural ideal amplitude is therefore
\begin{equation}
\Delta E(\theta)\equiv E(Y=0)-E(Y=1)\approx 1-\cos\theta.
\label{eq:deltaEideal}
\end{equation}

This architecture provides exactly the required operational isolation: the higher-order object is the context $Y$, while the measured probe observable remains strictly local.

\subsection{Concrete Kernel Quantification for A6}

To show that the higher-order context-conditioned kernel is not merely a label but a computable object, consider the ideal A6 parity ensemble with
\[
g_Y(S)\equiv \MI(Y;C_S),\qquad Y=C_0\oplus C_1\oplus C_2.
\]
By construction,
\[
\begin{aligned}
\MI(Y;C_i)&=0,\qquad \MI(Y;C_iC_j)=0,\\
\MI(Y;C_0C_1C_2)&=1~\text{bit},
\end{aligned}
\]
so M\"obius inversion on the subset lattice allocates the entire label-relevant constraint to the tripartite atom:
\begin{equation}
\begin{aligned}
f_Y(C_0,C_1,C_2)
&=\MI(Y;C_0C_1C_2)-\MI(Y;C_0C_1)\\
&\quad -\MI(Y;C_0C_2)\\
&\quad -\MI(Y;C_1C_2)+\MI(Y;C_0)\\
&\quad +\MI(Y;C_1)+\MI(Y;C_2)\\
&=1~\text{bit}.
\end{aligned}
\label{eq:a6mobius}
\end{equation}
Equivalently, within either parity-conditioned ensemble one may write the same tripartite atom in Shannon-entropy form as
\begin{align}
f_Y(C_0,C_1,C_2)
&=
S(C_0C_1\mid Y)+S(C_0C_2\mid Y)\nonumber\\
&\quad +S(C_1C_2\mid Y)\nonumber\\
&\quad -S(C_0\mid Y)-S(C_1\mid Y)\nonumber\\
&\quad -S(C_2\mid Y)-S(C_0C_1C_2\mid Y)\nonumber\\
&=1~\text{bit},
\label{eq:a6entropyform}
\end{align}
because each parity class is uniform over four allowed triples. In the idealized harness,
\[
\tilde f_Y^+(S)=0\quad (|S|<3),\qquad
\tilde f_Y^+(C_0,C_1,C_2)=1.
\]
Since the programmed context directly gates the conditional interaction, the operational overlap can be taken, to leading order, as
\[
\omega_v(S,\EXmean)\propto 1-\cos\theta.
\]
The residual kernel therefore reproduces the ideal A6 amplitude class,
\begin{equation}
\Gamma_{\rm rel}[Y,\EXmean]\propto (1~\text{bit})\times (1-\cos\theta),
\label{eq:a6kernel}
\end{equation}
which is the same functional dependence already observed in Eq.~(\ref{eq:deltaEideal}).

\subsection{Controls: parity balancing, passive nulls, and CTXONLY}

A6v23 strengthened the original v1 run in exactly the ways a skeptical hardware referee would ask for. First, it introduced parity-balancing replicates (R0/R1), which swap the even/odd labels across the same physical lanes and thereby neutralize the simple objection that one parity condition was mapped to uniformly better hardware lanes \cite{SramekA6Context2026}. The primary metric becomes the lane-balanced estimate
\begin{equation}
\Delta E_{\rm lane\mbox{-}balanced}(\theta)
=
\frac{1}{L}\sum_{\ell=1}^{L}
\big(E_{\ell,\mathrm{even}}(\theta)-E_{\ell,\mathrm{odd}}(\theta)\big).
\label{eq:lbalanced}
\end{equation}

Second, A6v23 included two passive controls:
\begin{itemize}
\item \textbf{PASSIVE1}: no parity compute and no conditional gate,
\item \textbf{PASSIVE2}: parity compute/uncompute but still no conditional gate,
\end{itemize}
thereby distinguishing ordinary lane drift from any artifact caused by the parity-computation scaffold itself \cite{SramekA6Context2026}.

Third, the run included \textbf{CTXONLY} circuits, used both for lane-SPAM screening and for checking that the intended HOIC-only context had not degraded into obvious low-order leakage. This is not window dressing; it is part of what allows the paper to use the phrase ``context-conditioned'' with a straight face.

\subsection{Measured results}

\paragraph*{Statistical methodology.}
All reported A6 empirical values are aggregated across the 8 physical lanes. Error bounds, visualized as standard errors of the mean (SEM), reflect inter-lane device heterogeneity, which empirically dominates simple binomial shot-noise contributions in this regime. Mutual-information estimates such as $\MI(Y;AB)$ are computed from empirical plug-in estimators, with Miller--Madow finite-sample bias correction applied at the entropy level before forming the corresponding information quantities. Confidence intervals for informational diagnostics are obtained via empirical bootstrapping over the shot records. To verify that the lane-balancing pipeline itself does not imprint spurious structure, we also use an ``ACTIVE randomized'' control in which the $Y$ labels are post-hoc relabeled at random within the same analysis pipeline. This defines the empirical null for the nonparametric label-permutation tests and centers $\Delta E$ at zero in expectation. The current archived artifacts report finite Monte Carlo resolutions set by the executed shuffle budgets, and the same workflow scales straightforwardly to future $10^5$-shuffle discovery-mode scans.

The v1 run on \texttt{ibm\_boston} used 8 lanes, 2048 shots per circuit, 8 circuits total, and 19 billed quantum-seconds. At $\theta=\pi/2$ in the XX basis it already showed a large split,
\[
\begin{aligned}
E_{\mathrm{even}} &= 0.878174,\qquad E_{\mathrm{odd}} = 0.00744629,\\
\Delta &= 0.870728,
\end{aligned}
\]
with $\MI(Y;AB)=0.200366$, triplet context mass $f_{C_0C_1C_2}=0.610731$, and a permutation-test $p$-value of $0.00497512$ \cite{SramekA6Context2026}. But v1 still left room for lane-bias and passive-drift objections.

\begin{figure*}[t]
\centering
\includegraphics[width=0.48\textwidth]{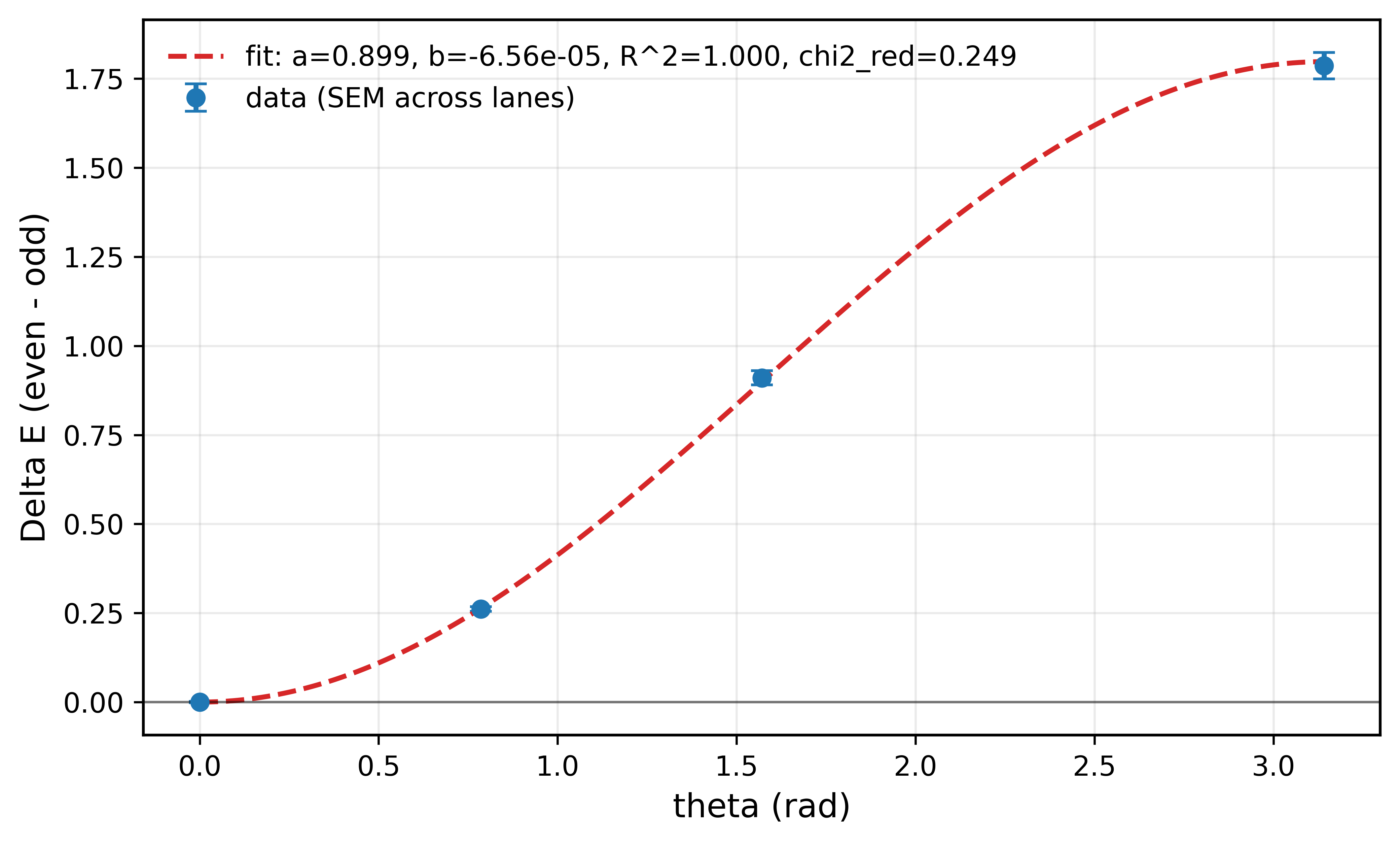}\hfill
\includegraphics[width=0.48\textwidth]{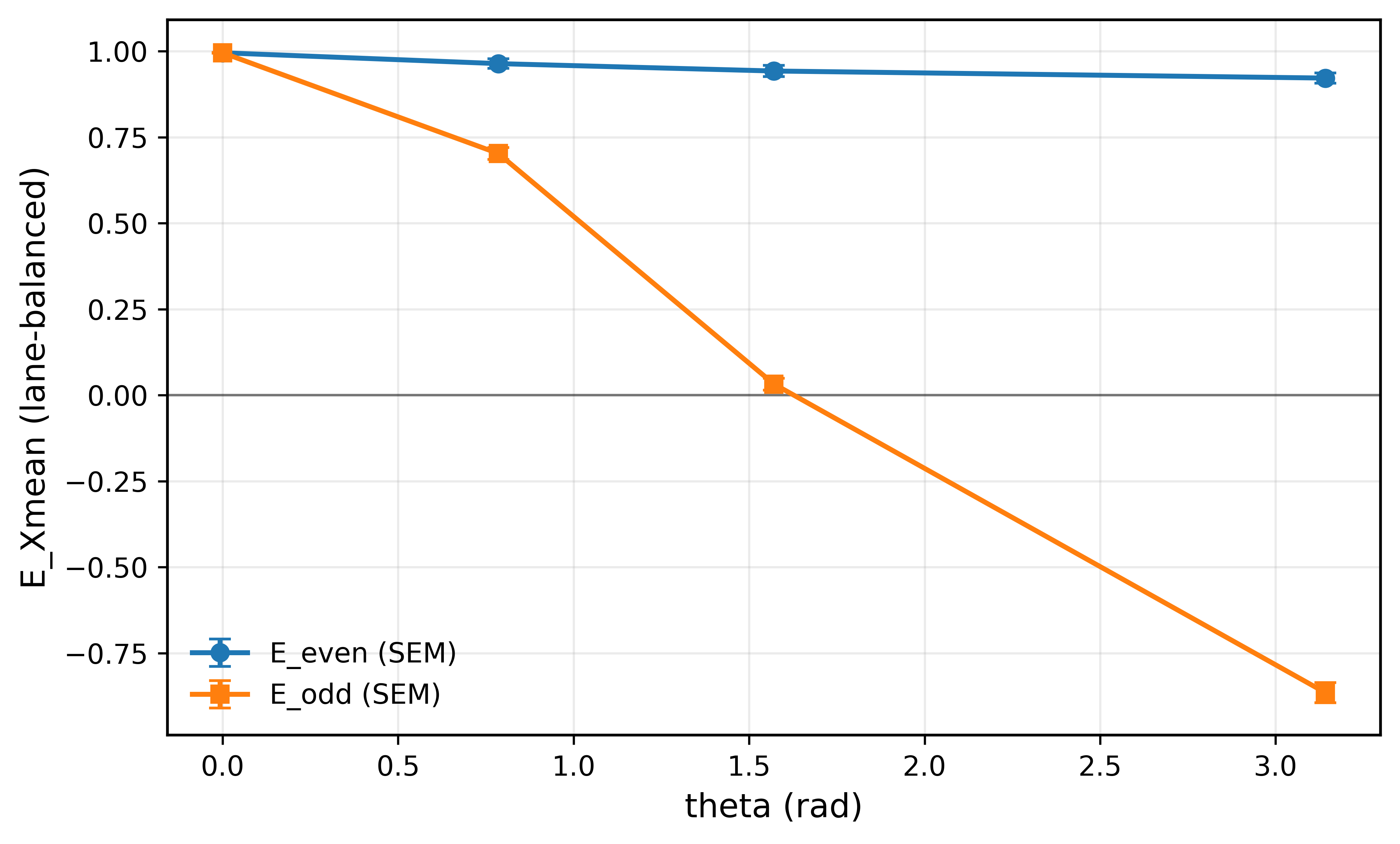}\\[0.6em]
\includegraphics[width=0.48\textwidth]{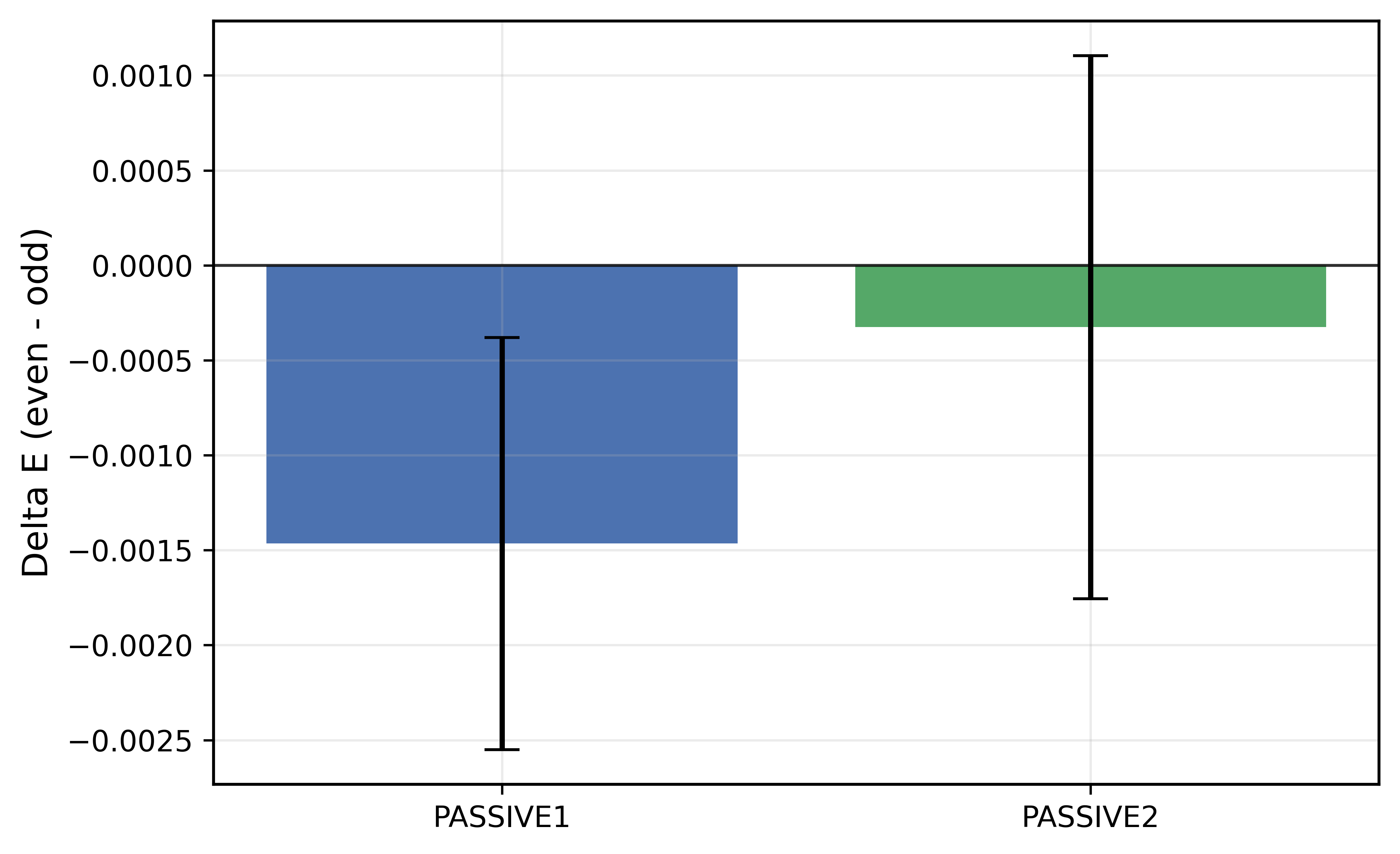}\hfill
\includegraphics[width=0.48\textwidth]{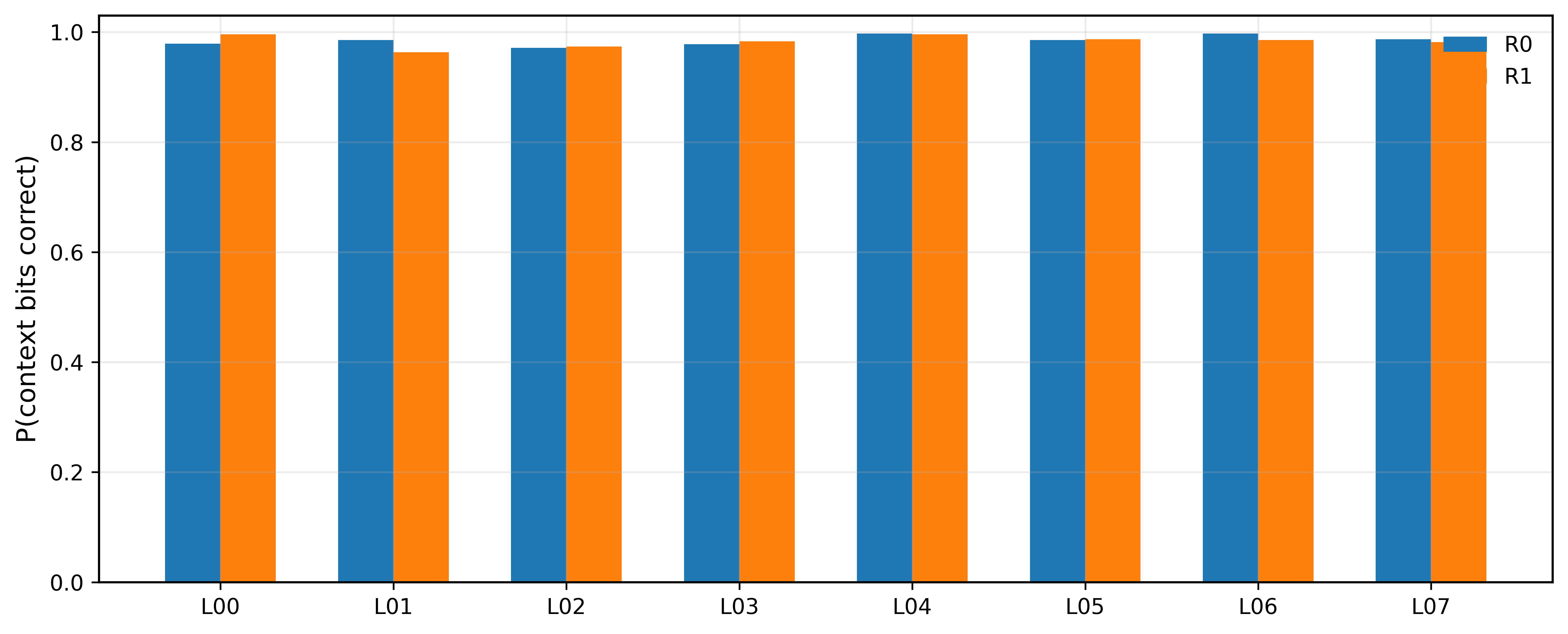}
\caption{\textbf{A6v23 hardware evidence for context-dependent local dynamics on \texttt{ibm\_boston}.}
Panel (a) shows the lane-balanced parity effect $\Delta E=E_{\mathrm{even}}-E_{\mathrm{odd}}$ in the XX basis as a function of the programmed interaction angle $\theta$. The data (blue points with SEM error bars across 8 lanes) are overlaid with a theoretical fit $\Delta E(\theta) = a(1-\cos\theta)+b$ (red dashed line), yielding an $R^2=1.000$ and a reduced $\chi^2 = 0.249$, confirming the exact expected functional form of the synthetic harness. Panel (b) resolves the corresponding lane-balanced branch values $E_{\mathrm{even}}$ and $E_{\mathrm{odd}}$. Panel (c) shows the passive controls PASSIVE1 and PASSIVE2; while PASSIVE1 shows a statistically resolvable drift, the SEM error bars strictly bound the magnitude to the order of $\sim 10^{-3}$, which is completely dwarfed by the active signal, proving the compute/uncompute scaffold alone does not induce the massive context-dependent effect. Panel (d) shows the CTXONLY lane-quality screen, confirming the cleanliness of the pairwise-matched parity context.\cite{SramekA6Context2026}}
\label{fig:a6summary}
\end{figure*}

The stronger evidence is the bundled v23 job: 8 lanes, 768 shots per circuit, 24 circuits total, and only 6 billed quantum-seconds on \texttt{ibm\_boston} \cite{SramekA6Context2026}. Its lane-balanced XX results are shown in Table~\ref{tab:a6active}, and the corresponding hardware plots are collected in Fig.~\ref{fig:a6summary}. The key features are exactly what one would want from a controlled harness: a near-zero baseline at $\theta=0$, a monotone increase with $\theta$, and a large response at $\theta=\pi/2$ and $\pi$.

\begin{table}[t]
\caption{\textbf{A6v23 lane-balanced ACTIVE results in the XX basis.} The observed response is near zero at $\theta=0$ and grows monotonically with the programmed conditional interaction angle, as expected for a controlled context-dependent dynamics harness \cite{SramekA6Context2026}.}
\label{tab:a6active}
\begin{ruledtabular}
\begin{tabular}{lccc}
$\theta$ & $E_{\mathrm{even}}$ & $E_{\mathrm{odd}}$ & $\Delta E$ \\
\hline
$0$       & $0.995443$  & $0.995443$  & $\sim 0$ \\
$\pi/4$   & $0.963867$  & $0.702637$  & $0.261230$ \\
$\pi/2$   & $0.942383$  & $0.0322266$ & $0.910156$ \\
$\pi$     & $0.921875$  & $-0.864909$ & $1.78678$ \\
\end{tabular}
\end{ruledtabular}
\end{table}

The passive controls remain statistically bounded near zero:
\[
\begin{aligned}
\Delta E_{\rm PASSIVE1} &= -0.00146484,\\
\Delta E_{\rm PASSIVE2} &= -0.000325521,
\end{aligned}
\]
again in the XX basis \cite{SramekA6Context2026}. CTXONLY screening reports worst-lane quality
\[
L01 = 0.963542
\]
for the minimum of $(\mathrm{ctx\_ok\_R0},\mathrm{ctx\_ok\_R1})$, which is good enough for the present regime and markedly cleaner than the early v1 concern about bad-lane contamination \cite{SramekA6Context2026}. At $\theta=\pi/2$, XX, ACTIVE R0, the observed mutual information was
\[
\MI(Y;AB)=0.259952
\]
with permutation-test $p=0.00199601$ \cite{SramekA6Context2026}.

\begin{table}[t]
\caption{\textbf{A6v23 passive controls in the XX basis.} These are the main confound killers: the parity-balancing analysis itself does not create a spurious signal, and parity compute/uncompute without the conditional $ZZ(\theta)$ interaction does not reproduce the ACTIVE effect \cite{SramekA6Context2026}.}
\label{tab:a6passive}
\begin{ruledtabular}
\begin{tabular}{lccc}
Family & $E_{\mathrm{even}}$ & $E_{\mathrm{odd}}$ & $\Delta E$ \\
\hline
PASSIVE1 & $0.995768$ & $0.997233$ & $-0.00146484$ \\
PASSIVE2 & $0.996745$ & $0.997070$ & $-0.000325521$ \\
\end{tabular}
\end{ruledtabular}
\end{table}

\subsection{Interpretation and scope}

What A6 establishes is strong but limited. It shows that a HOIC-pure context label can be engineered on current hardware, that the label can be screened to suppress low-order leakage, and that a local witness can nevertheless track that context with a large, dose-dependent response while passive controls remain near zero \cite{SramekA6Context2026}. In the language of Box~1, the proxy-only model is operationally incomplete for this harness.

What A6 does \emph{not} establish is that arbitrary natural interactions in the device are fundamentally DAGI-driven. The experiment is engineered. That is precisely why it is so useful: it validates a controlled template for later discovery-mode scans in which no explicit context-to-probe gate is present and any residual HOIC$\to$local coupling would have to emerge from the hardware itself \cite{SramekA6Context2026}. For the present paper, A6 is the primary evidence that the residual term $\Gamma_{\rm rel}[Y,\mathcal{O}]$ belongs in a phenomenological compression ansatz.

\section{Coherent Controllability: A6.2 as a Quantum-Eraser Extension}
\label{sec:a62}

\subsection{Coherent controllability and the quantum-eraser route}

To demonstrate that the context-dependent suppression observed in A6 is a programmable informational feature rather than irreversible hardware damage, we must establish coherent controllability. The cleanest route is an orthodox, multiqubit, hardware-native implementation of textbook quantum-eraser logic \cite{Scully1982,Kim2000,SramekA62Erasure2026}. The claim is not that an unconditional global witness revives after extra hardware depth. Rather, the claim is that MARK transfers coherence into system--marker correlations, and that eraser-basis conditioning restores large fringes in the conditional view.

\subsection{Minimal theory: MARK exports coherence into correlations}

Consider a system $S$ prepared in a coherent superposition of two effective histories,
\[
\lvert \psi\rangle_S = \alpha\lvert A\rangle + \beta\lvert B\rangle,
\]
and a marker ancilla $M$ that becomes correlated with the branch label. A programmable MARK interaction produces
\[
\lvert \Psi(\lambda)\rangle_{SM}
=
\alpha\lvert A\rangle\lvert m_A(\lambda)\rangle
+
\beta\lvert B\rangle\lvert m_B(\lambda)\rangle,
\]
where $\lambda$ is the marker-strength parameter \cite{SramekA62Erasure2026}. Tracing out the marker yields a reduced system state whose off-diagonal coherence is proportional to
\begin{equation}
\eta(\lambda)=\langle m_B(\lambda)\vert m_A(\lambda)\rangle.
\label{eq:eta}
\end{equation}
As the marker states become more distinguishable, $|\eta(\lambda)|$ decreases and the \emph{unconditional} interference visibility collapses. This is the precise sense in which MARK exports coherence into correlations rather than annihilating it \cite{SramekA62Erasure2026}.

This export of coherence is quantitatively bounded by the Englert--Greenberger--Yasin complementarity duality relation,
\begin{equation}
V^2 + \mathcal{D}^2 \le 1,
\label{eq:englert}
\end{equation}
where $V$ is the unconditional fringe visibility and $\mathcal{D}$ is the path distinguishability available from the marker $M$ \cite{Englert1996}. For the complementarity diagnostic in Fig.~\ref{fig:a62main}, $\mathcal{D}(\lambda)$ is estimated directly from the classical readout distributions as the total variation distance between the marker states conditioned on the complementary probe branches.

If the marker is then measured in an eraser basis $\{\lvert e_k\rangle\}$ that is not aligned with the which-path states, the conditional system state becomes
\[
\rho_{S|k}(\lambda)\propto
\langle e_k\vert \Psi(\lambda)\rangle
\langle \Psi(\lambda)\vert e_k\rangle,
\]
and the interference can reappear in the \emph{conditional} statistics \cite{Scully1982,Kim2000,SramekA62Erasure2026}. This is the textbook quantum-eraser mechanism, transplanted into a hardware-native multi-qubit circuit.

\subsection{A6.2 implementation and supported scope}

A6.2 was executed on IBM \texttt{ibm\_boston} (job \texttt{d5h1h92gim5s73ahd92g}) with 45 circuits, 384 shots per circuit, 9 billed quantum-seconds, and the physical motif
\[
q_1=50,\quad q_2=51,\quad q_3=52,\quad \mathrm{anc}=58
\]
\cite{SramekA62Erasure2026}. The experiment has three stages: interference preparation in a GHZ-like register, programmable MARK coupling between the register and a marker ancilla, and readout of the register plus eraser-basis analysis of the marker \cite{SramekA62Erasure2026}.

The \emph{supported} empirical claim in the current A6.2 draft is intentionally narrow. It is:
\begin{enumerate}
\item increasing MARK strength suppresses \emph{unconditional} fringes in the system marginal, and
\item conditioning on the marker measured in an eraser basis restores large fringes in the conditional view.
\end{enumerate}
The A6.2 source explicitly does \emph{not} yet claim a strong separation from an optimally matched local-dephasing control channel; that remains future work \cite{SramekA62Erasure2026}. We keep that scope control intact here.

\subsection{Results and interpretation}

\begin{figure*}[t]
\centering
\includegraphics[width=0.47\textwidth]{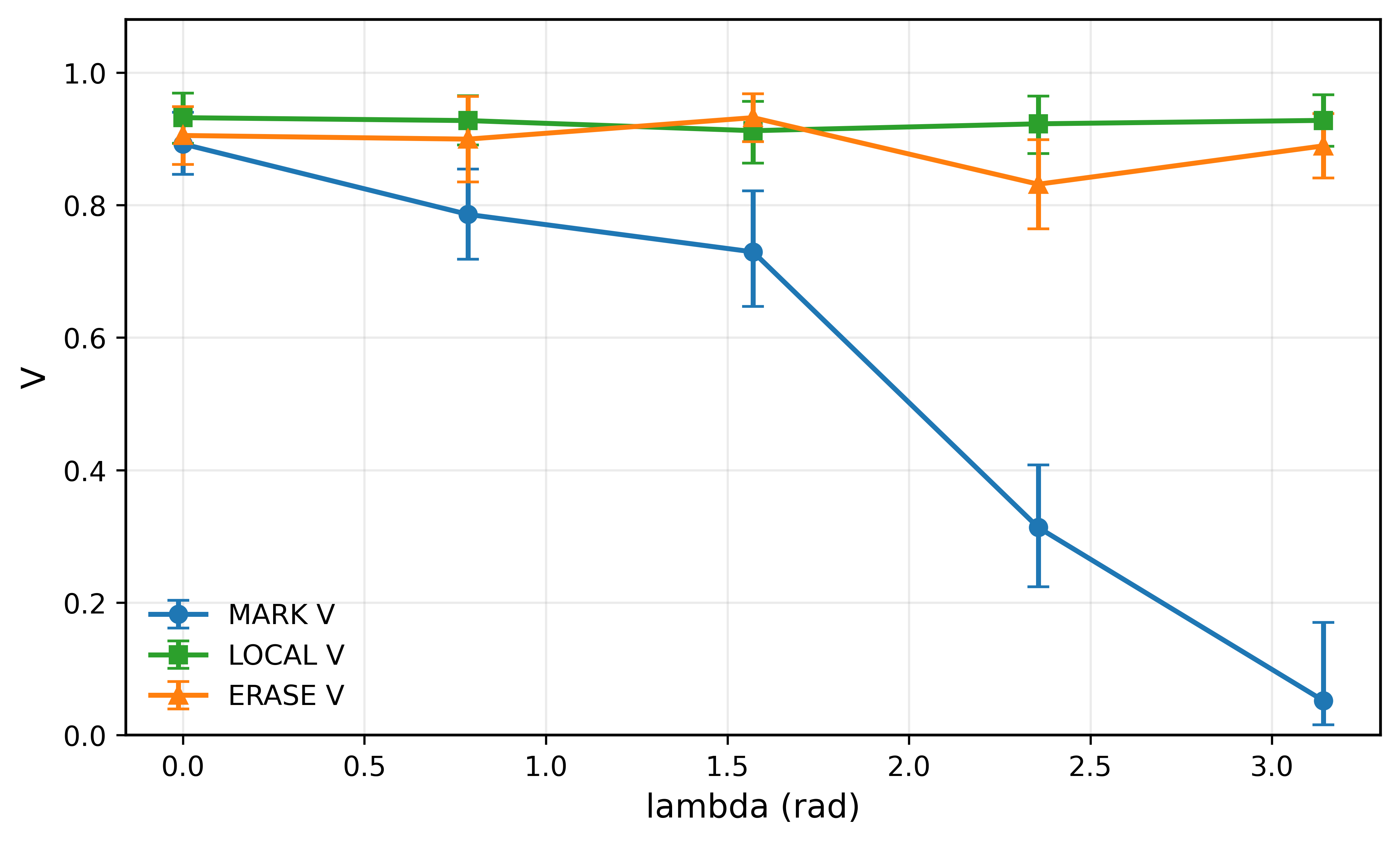}\hfill
\includegraphics[width=0.47\textwidth]{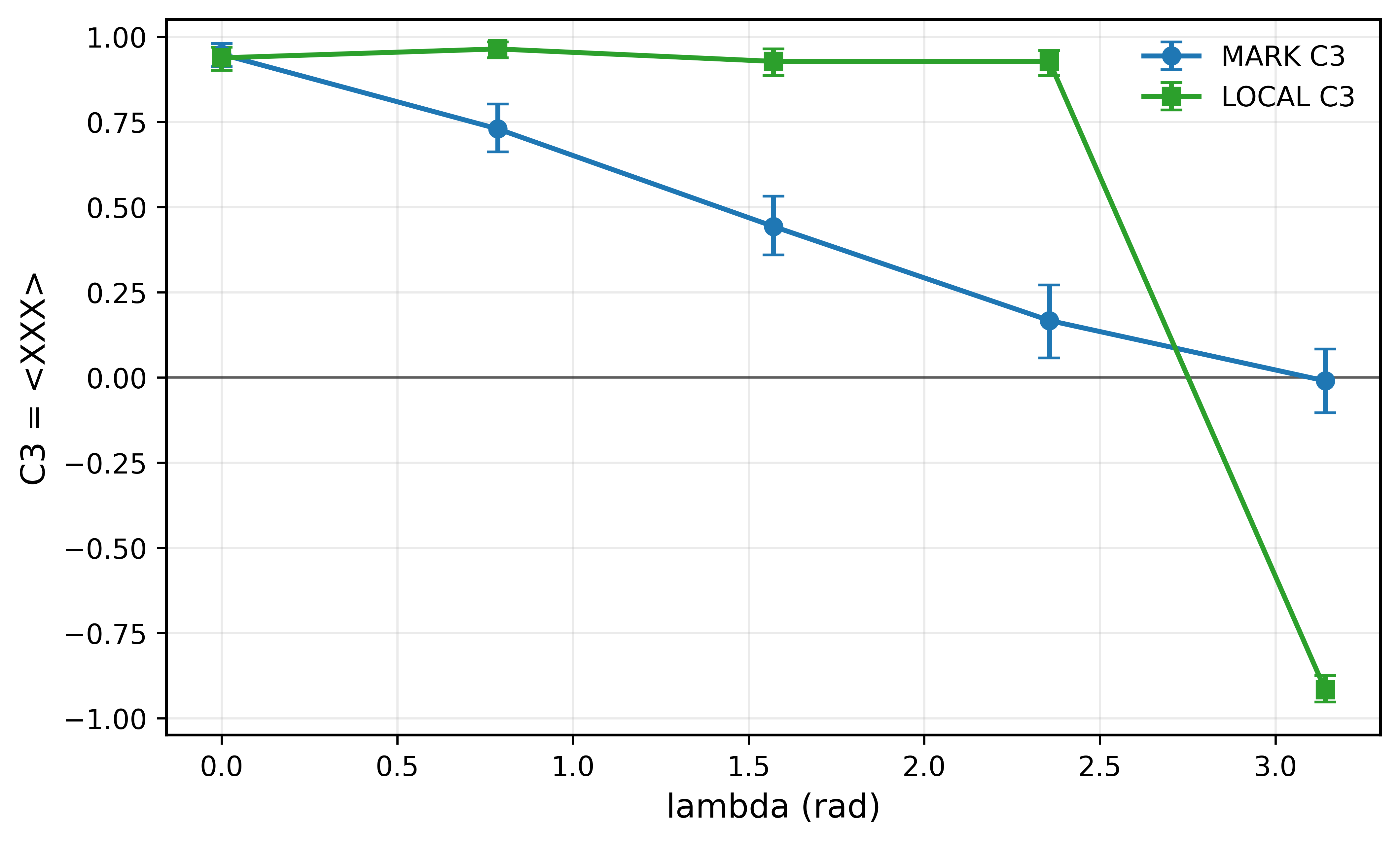}\\[0.45em]
\includegraphics[width=0.47\textwidth]{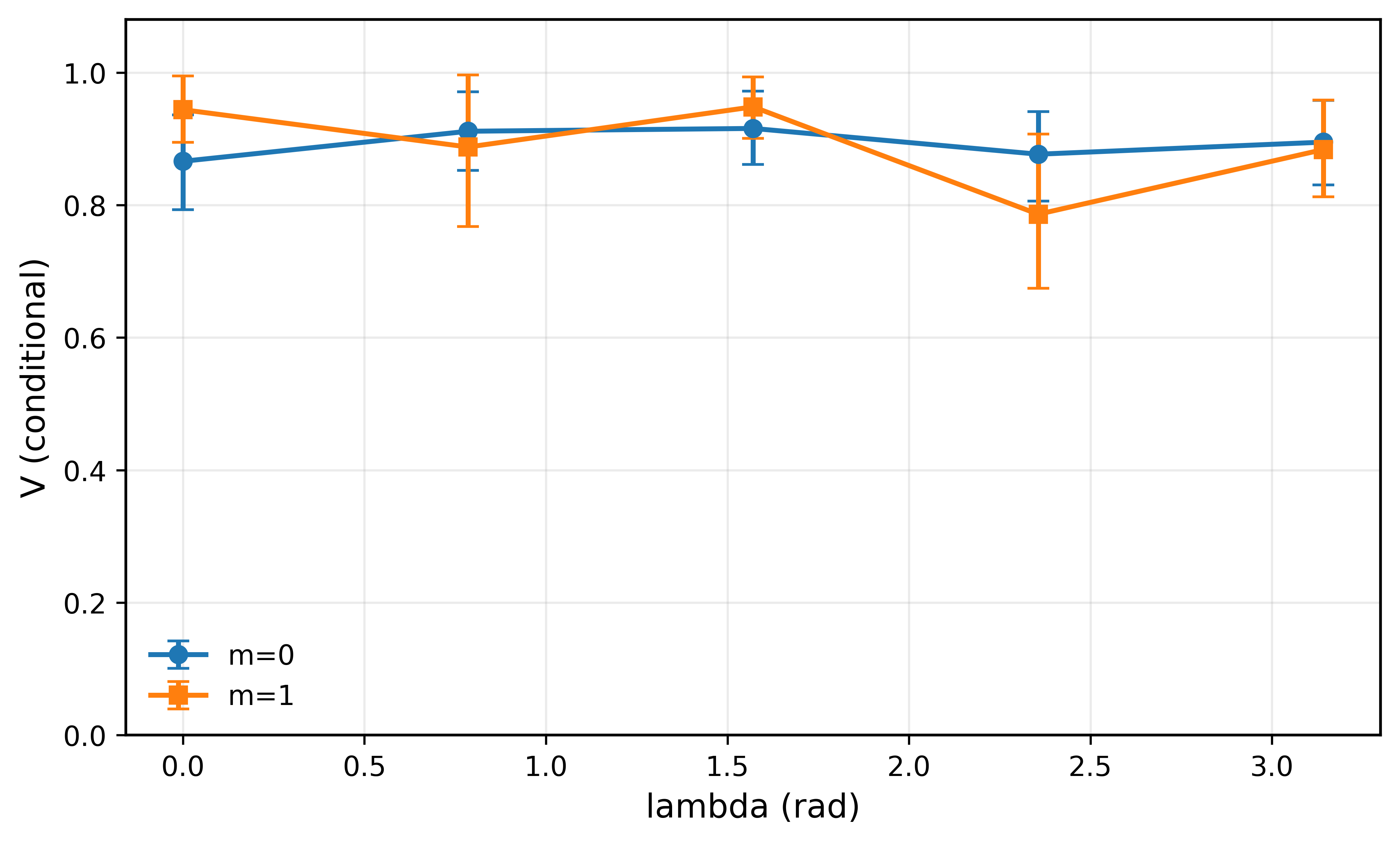}\hfill
\includegraphics[width=0.47\textwidth]{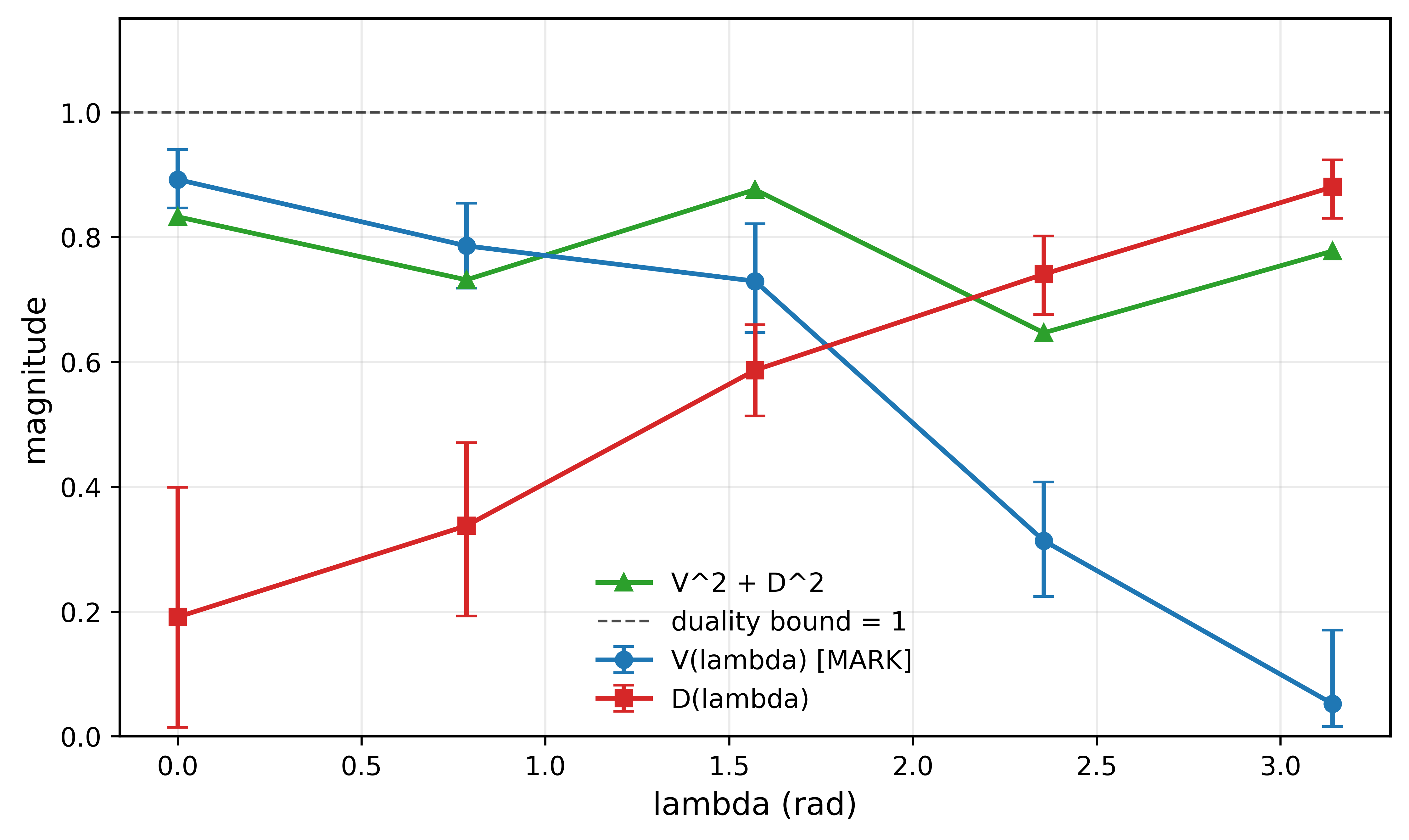}
\caption{\textbf{A6.2 MARK/ERASE results, tag diagnostics, and complementarity bounds on \texttt{ibm\_boston}.}
Panel (a) plots the probe-fringe magnitude $V(\lambda)$ versus marker strength $\lambda$, showing strong suppression of the unconditional marginal (MARK) while the conditional average (ERASE) remains large; the LOCAL curve is a hardware reference rather than an optimized local-dephasing null. Panel (b) tracks the global-tag observable $C_3(\lambda)=\langle XXX\rangle$ for the MARK branch and a local-matched reference as an auxiliary diagnostic of the applied marker interaction. Panel (c) resolves the conditional visibilities by eraser-basis outcome ($m=0,1$). Panel (d) tests the Englert--Greenberger--Yasin complementarity bound: as marker distinguishability $\mathcal{D}$ increases, unconditional visibility $V$ drops while the structural bound $V^2+\mathcal{D}^2\le 1$ is maintained across the sweep. The supported claim is the standard quantum-eraser one: MARK exports unconditional interference into system--marker correlations, while eraser-basis conditioning restores large fringes.\cite{Scully1982,Kim2000,SramekA62Erasure2026,Englert1996}}
\label{fig:a62main}
\end{figure*}

The A6.2 run shows the expected pattern, summarized in Fig.~\ref{fig:a62main}. As $\lambda$ is increased, the unconditional local fringe proxy $V_{\rm MARK}(\lambda)$ decreases strongly: near $\lambda=0$ it is close to its unmarked value, while near the strongest MARK point in the sweep it becomes very small \cite{SramekA62Erasure2026}. At the same time, conditioning on the marker outcome measured in an eraser basis yields large conditional visibilities $V_{\rm cond}(\lambda)$ across the sweep, including in the strong-MARK regime where the unconditional visibility is already suppressed \cite{SramekA62Erasure2026}.

A subtle but important point follows. A textbook quantum eraser is fundamentally about \emph{conditional} recovery; averaging over complementary conditioned fringes need not produce an unconditional revival. The relevant null hypothesis for A6.2 is therefore that the MARK interaction acts as an \emph{irreversible local-dephasing channel}---a standard Markovian error that permanently randomizes the phase of the probe. If the exported coherence were irreversibly lost to an unstructured bath, the ERASE conditional view would not be expected to show large, outcome-resolved fringes beyond the LOCAL reference. Because the MARK channel allows the ERASE subensembles to retain substantial phase information, the data strongly disfavors the irreversible local-dephasing null and is instead consistent with coherence being coherently exported into a retrievable system--marker correlation. At the same time, the analysis avoids any language implying that A6.2 proves unconditional macroscopic restoration or definitively separates from a pathologically optimized, non-Markovian local-dephasing control. That would require a separate, stringently matched dataset.

The arithmetic analogue sharpens the scope of this statement. In the squarefree divisor / log-prime model of Eqs.~(\ref{eq:Deltax})--(\ref{eq:softlimit}), the hard quota plays the role of a MARK-like boundary: it creates a discontinuous shell that strands isolated $k=1$ primes and limits parity-resolved cancellation to about $81.4\%$ in direct evaluations of the hard-cut model. Replacing that cliff by the softened Fermi--Dirac/$\tanh$ weight $W_{\tau,x}$ leaves the multiplicative graph itself unchanged, but continuously heals the shell, raises the mean cancellation efficiency to about $93.7\%$, and restores essentially perfect anti-correlation between the dominant parity sectors. The hardware and arithmetic settings should not be conflated: A6.2 does not claim unconditional macroscopic revival. What the arithmetic control shows is narrower and cleaner. In both settings, relaxing an artificially rigid boundary restores access to interference that was geometrically hidden by the hard partition.

\section{Discussion and Outlook}
\label{sec:discussion}

The framework presented here formalizes a strict operational target. First, the central mathematical object is not tied to the support size of the probe alone. The higher-order structure can live in the context $Y$ even when the measured witness is local. That matters because it lets the data attack a much harder target: not the familiar fragility of large GHZ states, but the steering of a local witness by a context label designed to be invisible to all low-order summaries of its carrier set.

Second, the proposed kernel is framed strictly as an operational tool. Equation~(\ref{eq:gammarel}) is a phenomenological compression ansatz, not a fully microscopic predictive theory. While Box~1 outlines the operational estimation pipeline utilizing classical inclusion--exclusion, full \emph{ab initio} prediction of $\omega_v$ across arbitrary circuits belongs to future work involving richer tomography and tighter null models. We acknowledge that standard low-order device proxies are not the only alternatives. Advanced frameworks such as quantum instrument linear gate set tomography (QILGST) \cite{Rudinger2022}, fast Bayesian tomography (FBT) \cite{Su2025FBT}, process-tensor tomography (PTT) \cite{Pollock2018,Milz2021}, and recent compressed multi-time learning methods such as temporal classical shadows \cite{White2024TemporalShadows} or open-quantum-evolution (OQE) learning from randomized benchmarking data \cite{Zhang2023OQE} can, in principle, reconstruct non-Markovian, context-dependent dynamics. However, these methods often carry steep experimental overheads, exponential scaling costs for full instrument reconstruction, or nontrivial matrix-product-operator classical post-processing that scales poorly for deep, multiscale algorithms. The $\Gamma_{\rm rel}$ kernel is not proposed as a replacement for full multi-time tomography, but as a lightweight, targeted diagnostic bridge. It compresses higher-order context dependence into a single measurable feature. This provides experimentalists with a scalable tool to rapidly flag regimes where mitigation techniques such as Measurement Randomized Compiling (MRC) \cite{Hashim2025MRC} fail to twirl away contextual crosstalk, allowing researchers to bound residual errors before committing to more expensive temporal-shadow or full process-tensor reconstructions.

The arithmetic model sharpens that point in the precise sense relevant to this paper. It removes the microwave stack, relaxation channels, and readout layer entirely. Equations~(\ref{eq:mertens-euler})--(\ref{eq:softlimit}) show that the same observable-conditioned hard-cut versus soft-cut logic can be written down exactly on a noiseless graph. In that setting, odd and even prime-factor layers contribute with alternating M\"obius signs to the Euler characteristic, the hard quota localizes topology into the shell $x/2 \le \prod p < x$, and the primes in $(x/2,x]$ are stranded as isolated 1-body vertices---the arithmetic counterpart of the isolated local clock. As our numerical evaluations show, the rigid boundary bottlenecks parity-cancellation efficiency near $81.4\%$; replacing the hard cliff by the softened logistic boundary (\texttt{ERASE}) raises the mean to about $93.7\%$ and drives the dominant parity sectors toward essentially perfect anti-correlation. This arithmetic control does not identify which physical channel realizes the effect on chip, and it should not be used to short-circuit proper QCVV. What it does eliminate is the narrower objection that the hard-cut versus soft-cut graph logic itself is a microwave artifact. The arithmetic construction therefore serves as a theorem-level control on the combinatorics of suppression and healing, while the separate empirical task remains to map those graph effects onto specific hardware mechanisms.

Third, the relation to relational physics can now be stated without overstretching. A6 and A6.2 are perfectly compatible with orthodox quantum theory. A6 is an engineered harness for higher-order context dependence. A6.2 is a implementation of the textbook quantum-eraser mechanism. What DAGI adds is a graph-structural interpretation of why record structure, coarse-graining, and accessible subsets matter for which histories can interfere. One may find that interpretation illuminating or unnecessary; either way, the operational content of the experiments stands on its own.

The immediate next steps are technical, not cosmological. The most important are: (i) scaling comparisons across context order at matched entropy and matched local depth, (ii) matched local-dephasing controls for the A6.2 MARK channel, (iii) explicit dynamical-decoupling studies during the relevant record windows, and (iv) discovery-mode runs with no explicit context-to-probe gate, where any non-null result would indicate genuine device-level HOIC sensitivity \cite{SramekA6Context2026,SramekA62Erasure2026}. A restrained long-range outlook is still possible: if irreversible record load grows under coarse-graining, then the available capacity for maintaining higher-order coupling structures may become correspondingly constrained. But the paper deliberately stops at that theorem-like hint and leaves all stronger cosmological extrapolations for later work.

\section{Conclusion}
\label{sec:conclusion}

Current device-level proxy metrics are often useful, but they are not always sufficient summaries of dynamic-circuit backaction. By introducing a higher-order context-conditioned kernel, we provide an operational framework to capture disturbance that strictly depends on the intersection between macroscopic context and queried observables. 

In the tested A6 harness, a context label deliberately hidden from all 1- and 2-body summaries of its carrier set nevertheless steered a local witness with a large, dose-dependent response while passive controls remained bounded near zero. In A6.2, programmable marker coupling suppressed unconditional interference while eraser-basis conditioning restored large conditional fringes. Together, these results strongly favor a context-conditioned description of backaction over a proxy-only baseline.

Whether this framework matures into a quantitatively predictive characterization tool will depend on the next round of hardware scaling, mitigation, and tomography. For now, its value is operational: it gives dynamic-circuit experiments a compact language for separating ordinary local/proxy noise from higher-order context effects, without pretending that the latter have already been solved from first principles.

\begin{acknowledgments}
The author acknowledges IBM Quantum for access to superconducting hardware used throughout the DAGI empirical validation program. The author also thanks the broader literature on dynamic-circuit characterization and decoherence theory for providing the operational baseline against which the present tests are framed.
\end{acknowledgments}

\section*{Statements and Declarations}

\paragraph*{Funding} The author did not receive any specific grant or funding for this work.

\paragraph*{Author contributions} Petr Sramek conceived the study; designed the theoretical framework and hardware harnesses; executed or supervised the circuit construction, data acquisition, and statistical analysis; and wrote the manuscript. The author read and approved the final manuscript.

\paragraph*{Competing interests} Petr Sramek is affiliated with Whytics and the DAGI Research Program. Beyond these affiliations, the author declares no competing financial or non-financial interests directly or indirectly related to the work presented in this manuscript.

\paragraph*{Ethics approval and consent to participate} Not applicable.

\paragraph*{Consent for publication} Not applicable.

\paragraph*{Data and code availability} The raw Qiskit schedules, circuit-generation scripts, analysis notebooks, plotting code, and supporting artifacts for the A6, A6.2, and related DAGI hardware studies are archived through the cited Zenodo records \cite{SramekA6Context2026,SramekA62Erasure2026,SramekWignersFriend2026,SramekTimeDilation2026,SramekRecordErasure2026}. These public records provide the minimal datasets and executable artifacts needed to inspect and reproduce the analyses reported here.

\appendix

\section{Auxiliary coherent diagnostic already embedded in A6v23}
\label{app:a6coherent}

The bundled A6v23 job already contained a compact coherent branch comparing \texttt{WHICH\_Z} and \texttt{ERASE\_X}. We do not use this as the primary coherent-evidence figure in the main text because the dedicated A6.2 sweep is cleaner and more extensive. It is nevertheless a useful internal consistency check: the \texttt{WHICH\_Z} branch yields classical branch separation, whereas \texttt{ERASE\_X} yields opposite-signed conditioned branches consistent with erasure logic on the same hardware motif.\cite{SramekA6Context2026}


\begin{thebibliography}{99}


\bibitem{Pollock2018}
F.~A.~Pollock, C.~Rodr\'iguez-Rosario, T.~Frauenheim, M.~Paternostro, and K.~Modi,
\emph{Operational Markov condition for quantum processes},
\href{https://doi.org/10.1103/PhysRevLett.120.040405}{Phys.\ Rev.\ Lett.\ \textbf{120}, 040405 (2018)}.

\bibitem{Milz2021}
S.~Milz and K.~Modi,
\emph{Quantum stochastic processes and quantum non-Markovian phenomena},
\href{https://doi.org/10.1103/PRXQuantum.2.030201}{PRX Quantum \textbf{2}, 030201 (2021)}.

\bibitem{Rudinger2022}
K.~Rudinger, G.~J.~Ribeill, L.~C.~G.~Govia, M.~Ware, E.~Nielsen, K.~Young, T.~A.~Ohki, R.~Blume-Kohout, and T.~Proctor,
\emph{Characterizing midcircuit measurements on a superconducting qubit using gate set tomography},
\href{https://doi.org/10.1103/PhysRevApplied.17.014014}{Phys.\ Rev.\ Applied \textbf{17}, 014014 (2022)}.

\bibitem{Govia2023}
L.~C.~G.~Govia, P.~Jurcevic, C.~J.~Wood, N.~Kanazawa, S.~T.~Merkel, and D.~C.~McKay,
\emph{A randomized benchmarking suite for mid-circuit measurements},
\href{https://doi.org/10.1088/1367-2630/ad0eda}{New J.\ Phys.\ \textbf{25}, 123016 (2023)}.

\bibitem{Hothem2025}
D.~Hothem, J.~Hines, C.~Baldwin, D.~Gresh, R.~Blume-Kohout, and T.~Proctor,
\emph{Measuring error rates of mid-circuit measurements},
\href{https://doi.org/10.1038/s41467-024-54350-0}{Nat.\ Commun.\ \textbf{16}, 5761 (2025)}. 

\bibitem{Scully1982}
M.~O.~Scully and K.~Dr\"uhl,
\emph{Quantum eraser: A proposed photon correlation experiment concerning observation and ``delayed choice'' in quantum mechanics},
\href{https://doi.org/10.1103/PhysRevA.25.2208}{Phys.\ Rev.\ A \textbf{25}, 2208--2213 (1982)}.

\bibitem{Kim2000}
Y.-H.~Kim, R.~Yu, S.~P.~Kulik, Y.~Shih, and M.~O.~Scully,
\emph{A delayed choice quantum eraser},
\href{https://doi.org/10.1103/PhysRevLett.84.1}{Phys.\ Rev.\ Lett.\ \textbf{84}, 1--5 (2000)}.

\bibitem{Zurek2003}
W.~H.~Zurek,
\emph{Decoherence, einselection, and the quantum origins of the classical},
\href{https://doi.org/10.1103/RevModPhys.75.715}{Rev.\ Mod.\ Phys.\ \textbf{75}, 715--775 (2003)}.

\bibitem{Schlosshauer2007}
M.~Schlosshauer,
\emph{Decoherence and the Quantum-To-Classical Transition}
(Springer, Berlin, 2007). \url{https://doi.org/10.1007/978-3-540-35775-9}

\bibitem{Rovelli1996}
C.~Rovelli,
\emph{Relational quantum mechanics},
\href{https://doi.org/10.1007/BF02302398}{Int.\ J.\ Theor.\ Phys.\ \textbf{35}, 1637--1678 (1996)}.

\bibitem{Giacomini2019}
F.~Giacomini, E.~Castro-Ruiz, and \v{C}.~Brukner,
\emph{Quantum mechanics and the covariance of physical laws in quantum reference frames},
\href{https://doi.org/10.1038/s41467-018-08155-0}{Nat.\ Commun.\ \textbf{10}, 494 (2019)}.

\bibitem{SramekA6Context2026}
P.~Sramek,
\emph{A6: Context-Dependent Local Dynamics from Pure Higher-Order Context on IBM Quantum Hardware: A Controlled DAGI Downward-Causation Harness with Passive Nulls and a Built-In Quantum Eraser Variant} (Draft v0.9),
Zenodo (2026). \url{https://doi.org/10.5281/zenodo.18911023}.

\bibitem{SramekA62Erasure2026}
P.~Sramek,
\emph{A6.2: Programmable Global Which-Path Tagging and Conditional Interference Restoration on Superconducting Quantum Hardware} (Draft v0.9),
Zenodo (2026). \url{https://doi.org/10.5281/zenodo.18911100}.

\bibitem{SramekWignersFriend2026}
P.~Sramek,
\emph{Observable-Selective Branching: A Hardware-Validated Topological Resolution to the Wigner's Friend Paradox} (v1.0),
Zenodo (2026). \url{https://doi.org/10.5281/zenodo.18921630}.

\bibitem{SramekTimeDilation2026}
P.~Sramek,
\emph{Informational Time Dilation on a Superconducting Quantum Processor: Schedule-Matched Evidence for Irreversibility-Controlled Clock Slowdown} (v1.0),
Zenodo (2026). \url{https://doi.org/10.5281/zenodo.18909441}.

\bibitem{SramekRecordErasure2026}
P.~Sramek,
\emph{Record, Erasure, and Distributed Witness Sensitivity on IBM Quantum Hardware: A DAGI Validation Study} (v1.0),
Zenodo (2026). \url{https://doi.org/10.5281/zenodo.18911684}.


\bibitem{Englert1996}
B.-G.~Englert,
\emph{Fringe Visibility and Which-Way Information: An Inequality},
\href{https://doi.org/10.1103/PhysRevLett.77.2154}{Phys.\ Rev.\ Lett.\ \textbf{77}, 2154--2157 (1996)}.

\bibitem{Su2025FBT}
R.~Y.~Su, J.~Y.~Huang, N.~D.~Stuyck, M.~K.~Feng, W.~Gilbert, T.~J.~Evans, W.~H.~Lim, F.~E.~Hudson, K.~W.~Chan, W.~Huang, K.~M.~Itoh, R.~Harper, S.~D.~Bartlett, C.~H.~Yang, A.~Laucht, A.~Saraiva, T.~Tanttu, and A.~S.~Dzurak,
\emph{Characterizing non-Markovian quantum processes by fast Bayesian tomography},
\href{https://doi.org/10.1103/PhysRevA.111.052425}{Phys.\ Rev.\ A \textbf{111}, 052425 (2025)}.

\bibitem{White2024TemporalShadows}
G.~A.~L.~White, L.~C.~L.~Hollenberg, C.~D.~Hill, and K.~Modi,
\emph{Practical learning of multi-time statistics in open quantum systems},
\href{https://arxiv.org/abs/2412.17862}{arXiv:2412.17862} (2024).

\bibitem{Zhang2023OQE}
X.~Zhang, Z.~Wu, G.~A.~L.~White, Z.~Xiang, S.~Hu, Z.~Peng, Y.~Liu, D.~Zheng, X.~Fu, A.~Huang, D.~Poletti, K.~Modi, J.~Wu, M.~Deng, and C.~Guo,
\emph{Randomised benchmarking for characterizing and forecasting correlated processes},
\href{https://arxiv.org/abs/2312.06062}{arXiv:2312.06062} (2023).

\bibitem{Hashim2025MRC}
A.~Hashim, A.~Carignan-Dugas, L.~Chen, C.~Juenger, N.~Fruitwala, Y.~Xu, G.~Huang, J.~J.~Wallman, and I.~Siddiqi,
\emph{Quasiprobabilistic readout correction of mid-circuit measurements for adaptive feedback via measurement randomized compiling},
\href{https://doi.org/10.1103/PRXQuantum.6.010307}{PRX Quantum \textbf{6}, 010307 (2025)}.

\bibitem{Bjorner2011}
A.~Bj\"orner,
\emph{A cell complex in number theory},
\href{https://doi.org/10.1016/j.aam.2010.09.007}{Adv.\ Appl.\ Math.\ \textbf{46}, 71--85 (2011)}.

\bibitem{Pakianathan2013}
J.~Pakianathan and T.~Winfree,
\emph{Threshold complexes and connections to number theory},
\href{https://doi.org/10.3906/mat-1112-14}{Turkish J.\ Math.\ \textbf{37}, 511--539 (2013)}.

\end{thebibliography}
\end{document}